\documentclass[runningheads]{llncs}
\usepackage[T1]{fontenc}
\usepackage{cite}
\usepackage{amsmath,amssymb,amsfonts}
\usepackage{graphicx}
\usepackage{textcomp}
\usepackage{xspace}
\usepackage{booktabs}
\usepackage{multirow}
\usepackage[ruled,vlined,linesnumbered]{algorithm2e} 
\usepackage{array}
\usepackage{balance}
\usepackage{color}
\usepackage{colortbl}
\usepackage{courier}
\usepackage{csvsimple}
\usepackage{enumitem}
\usepackage{fancybox}
\usepackage{listings}
\usepackage{longtable}
\usepackage{lscape}
\usepackage{makecell}
\usepackage{marvosym}
\usepackage{moreverb}
\usepackage{multicol}
\usepackage{pifont} 
\usepackage{rotating}
\usepackage{setspace}
\usepackage{subfigure}
\usepackage[most]{tcolorbox}
\usepackage{threeparttable}
\usepackage{tikz}
\usepackage{soul}
\usepackage[normalem]{ulem}
\usepackage{url}
\usepackage{wasysym}
\usepackage[numbers]{natbib} 
\usepackage{hyperref}
\usepackage{todonotes} 

\begin{document}
\title{Contrastive Augmentation: An Unsupervised Learning Approach for Keyword Spotting in Speech Technology}
%
%
\author{Weinan Dai\inst{1}\thanks{Weinan Dai and Yifeng Jiang contributed equally to this work.}\orcidID{0009-0004-1201-3471} \and
Yifeng Jiang\inst{2}\orcidID{0009-0008-9534-0185} \and
Yuanjing Liu\inst{3}\orcidID{0009-0002-9636-2526} \and
Jinkun Chen\inst{4}\orcidID{0009-0004-5792-2097} \and
Xin Sun\inst{5}\orcidID{0009-0006-4333-3990} \and
Jinglei Tao\inst{6}\orcidID{0009-0001-4709-4302}}
\authorrunning{W. Dai and Y. Jiang et al.}
\institute{Trine University, Phoenix, USA \\
\email{wnd17460@gmail.com} \and
Boston University, Boston, USA \\
\email{yjiang8@bu.edu} \and
Georgia Institute of Technology, Atlanta, GA, USA \\
\email{yliu3689@gatech.edu} \and
Faculty of Computer Sciences, Dalhousie University, Halifax, Canada \\
\email{jinkun.chen@dal.ca} \and
Texas A\&M University, College Station, TX, USA \\
\email{yolosun4488@gmail.com} \and
Georgia Institute of Technology, Atlanta, GA, USA \\
\email{jinglei.tao@gatech.edu}}

\maketitle              
\begin{abstract}
This paper addresses the persistent challenge in Keyword Spotting (KWS), a fundamental component in speech technology, regarding the acquisition of substantial labeled data for training. Given the difficulty in obtaining large quantities of positive samples and the laborious process of collecting new target samples when the keyword changes, we introduce a novel approach combining unsupervised contrastive learning and a unique augmentation-based technique. Our method allows the neural network to train on unlabeled data sets, potentially improving performance in downstream tasks with limited labeled data sets. We also propose that similar high-level feature representations should be employed for speech utterances with the same keyword despite variations in speed or volume. To achieve this, we present a speech augmentation-based unsupervised learning method that utilizes the similarity between the bottleneck layer feature and the audio reconstructing information for auxiliary training. Furthermore, we propose a compressed convolutional architecture to address potential redundancy and non-informative information in KWS tasks, enabling the model to simultaneously learn local features and focus on long-term information. This method achieves strong performance on the Google Speech Commands V2 Dataset. Inspired by recent advancements in sign spotting and spoken term detection, our method underlines the potential of our contrastive learning approach in KWS and the advantages of Query-by-Example Spoken Term Detection strategies. The presented CAB-KWS provide new perspectives in the field of KWS, demonstrating effective ways to reduce data collection efforts and increase the system's robustness.
\keywords{key word spotting \and contrastive learning \and unsupervised learning}
\end{abstract}
\section{INTRODUCTION}
Keyword Spotting (KWS) is a fundamental application in the field of speech technology, playing a pivotal role in real-world scenarios, particularly in the context of interactive agents such as virtual assistants and voice-controlled devices. KWS is designed to detect a small set of pre-defined keywords within an audio stream. This capability is crucial for two primary reasons. First, it enables the initiation of interactions through specific commands like "hey Siri" or "OK, Google," effectively serving as an explicit cue for the system to start processing subsequent speech. Second, KWS can identify sensitive words within a conversation, thereby playing a vital role in protecting the privacy of the speaker. Given these applications, it is crucial to develop accurate and reliable KWS systems for effective real-world speech processing~\cite{li2017acoustic,luo2021end,schalkwyk2010your}.

Despite the considerable advancements in KWS, a significant challenge that persists is the acquisition of sufficient labeled data for training. This is especially true for positive samples, which are often harder to obtain in large quantities. This issue is further exacerbated when the keyword changes, as it necessitates the collection of new target samples, a process that can be both time-consuming and resource-intensive. To address these challenges, we propose a novel approach that leverages the power of unsupervised contrastive learning and a unique augmentation-based method. Additionally, another potential problem is redundant information, speeches are noisy and complex, where only some key phrases are highly related to the keywords. However, convolutional methods treat all the word windows equally, ignoring that different words have different importance and should be weighted differently within word windows. Besides, the sliding windows used in the convolutional methods produce a lot of redundant information. Thus, it is important to reduce the non-informative and redundant information and distinguish the contributions of different convolutional features.

Our method enables the neural network to be trained on unlabeled datasets, reducing the reliance on extensive labeled data. This technique can greatly enhance the performance of downstream tasks, even in scenarios where labeled datasets are scarce. Additionally, we propose that speech utterances containing the same keyword, regardless of variations in speed or volume, should exhibit similar high-level feature representations in KWS tasks. To achieve this, we present a speech augmentation-based unsupervised learning approach. This method leverages the similarity of bottleneck layer features, along with audio reconstruction information, for auxiliary training to improve system robustness.

In addition to these innovations, we propose a compressed convolutional architecture for the KWS task. This architecture, designed to tackle the issue of redundant information, has demonstrated strong performance on the Google Speech Commands V2 Dataset. By doing so, it enables the model to learn local features and focus on long-term information simultaneously, thereby enhancing its performance on the KWS task.

Our approach is inspired by recent advancements in the field of sign spotting and spoken term detection. For instance, Varol et al.\cite{varol2022scaling} demonstrated the effectiveness of Noise Contrastive Estimation and Multiple Instance Learning in sign spotting, which could provide insights into the use of contrastive learning in KWS. Similarly, the works of Tejedor et al.\cite{tejedor2019search, tejedor2018albayzin} on Query-by-Example Spoken Term Detection (QbE STD) highlight the potential of QbE STD strategies in outperforming text-based STD in unseen data domains, reinforcing the potential advantages of our proposed method.

Our major contributions in this work are as follows:

\begin{itemize}
\item We introduce a compact convolutional architecture for the KWS task that achieves strong results on the Google Speech Commands V2 Dataset.
\item We develop an unsupervised loss and a contrastive loss to evaluate the similarity between original and augmented speech, as well as the proximity within each minibatch.
\item We introduce a speech augmentation-based unsupervised learning approach, utilizing the similarity between the bottleneck layer feature, as well as the audio reconstructing information for auxiliary training.
\end{itemize}

Theremainder of this paper is structured as follows. Section~\ref{sec:relatedwork} provides an overview of related work in the areas of data augmentation, unsupervised learning, and other methodologies of KWS tasks. Section~\ref{sec:preli} offers a background on contrastive learning. Section~\ref{sec:approach} details the proposed model architecture and our augmentation-based unsupervised contrastive learning loss. Section~\ref{sec:setup} discusses the configuration, research questions, and experimental setups. Section~\ref{sec:results} presents the experimental results and compares them with other pre-training methods. We also discuss the relationship between pre-training steps and the performance of downstream KWS tasks. Finally, Section~\ref{sec:conclusion} concludes the paper with a summary of our findings and potential avenues for future work.

\section{RELATED WORK} \label{sec:relatedwork}

Data augmentation is widely acknowledged as an effective technique for enriching the training datasets in speech applications, such as Automatic Speech Recognition (ASR) and Keyword Spotting (KWS). Various methods have been explored, such as vocal tract length perturbation~\cite{jaitly2013vocal}, speed-perturbation~\cite{ko2015audio}, and the introduction of noisy audio signals~\cite{hannun2014deep}. More recently, spectral-domain augmentation techniques, such as SpecAugment~\cite{park2019specaugment} and WavAugment~\citep{kharitonov2021data}, have been developed to further improve the robustness of speech recognition systems. In this work, we extend these efforts by applying speed and volume perturbation in our speech augmentation method.

While supervised learning has been the primary approach in the KWS area, it often requires large amounts of labeled data, which can be challenging to obtain, especially for less frequently used languages. This has sparked growing interest in weakly supervised and unsupervised approaches. For example, Noisy Student Training, a semi-supervised learning technique, has been employed in ASR~\cite{park2020improved} and subsequently adapted for robust keyword spotting~\citep{park2021noisy}. Additionally, unsupervised methods for KWS have been investigated~\citep{garcia2006keyword,li2007novel,zhang2009unsupervised}, yielding promising outcomes. Building on these efforts, we propose an unsupervised learning framework for the keyword spotting task in this paper.

The Google Speech Commands V2 Dataset is a widely used benchmark for novel ideas in KWS. Numerous works have performed experiments on this dataset, introducing various architectures and methods. For instance, a convolutional recurrent network with attention was introduced by \citep{de2018neural}, and a deep residual network, MatchboxNet, was proposed by \citep{majumdar2020matchboxnet}. More recently, an edge computing-focused model called EdgeCRNN~\citep{wei2022edgecrnn} was introduced, along with a method that integrates triplet loss-based embeddings with a modified K-Nearest Neighbor (KNN) for classification~\citep{vygon2021learning}. In this work, we also evaluate our speech augmentation-based unsupervised learning method on this dataset and compare it with other unsupervised approaches, including CPC~\citep{oord2018representation}, APC~\citep{chung2019unsupervised}, and MPC ~\citep{jiang2019improving}.

\section{PRELIMINARY STUDY OF CONTRASTIVE LEARNING} \label{sec:preli}
In the context of a classification task involving $K$ classes, we consider a dataset $\left\{\boldsymbol{x}_{i}, y_{i}\right\}_{i=1}^{N}$ with $N$ training samples. Each $\boldsymbol{x}{i} \in \mathbb{R}^{L}$ represents an input sentence of $L$ words, and each $y_{i} \in{1,2, \cdots, K}$ is the corresponding label. We denote the set of training sample indexes by $\mathcal{I}={1,2, \cdots, N}$ and the set of label indexes by $\mathcal{K}={1,2, \cdots, K}$.

We explore the realm of self-supervised contrastive learning, a technique that has demonstrated its effectiveness in numerous studies. Given $N$ training samples $\left\{\boldsymbol{x}_{i}\right\}_{i=1}^{N}$ with a number of augmented samples, the standard contrastive loss is defined as follows:

\begin{equation}
\mathcal{L}{\text {self }}=\frac{1}{N} \sum{i \in \mathcal{I}}-\log \frac{\exp \left(\boldsymbol{z}{i} \cdot \boldsymbol{z}{j(i)} / \tau\right)}{\sum_{a \in \mathcal{A}{i}} \exp \left(\boldsymbol{z}{i} \cdot \boldsymbol{z}_{a} / \tau\right)}
\end{equation}

Here, $\boldsymbol{z}{i}$ is the normalized representation of $\boldsymbol{x}{i}, \mathcal{A}_{i}:=\mathcal{I} \backslash{i}$ is the set of indexes of the contrastive samples, the $\cdot$ symbol denotes the dot product, and $\tau \in \mathbb{R}^{+}$ is the temperature factor.

However, self-supervised contrastive learning does not utilize supervised signals. A previous study [Khosla et al., 2020] incorporated supervision into contrastive learning in a straightforward manner. It simply treated samples from the same class as positive samples and samples from different classes as negative samples. The following contrastive loss is defined for supervised tasks:

\begin{equation}
\mathcal{L}{\text {sup }}=\frac{1}{N} \sum{i \in \mathcal{I}} \frac{1}{\left|\mathcal{P}{i}\right|} \sum{p \in \mathcal{P}{i}}-\log \frac{\exp \left(\boldsymbol{z}{i} \cdot \boldsymbol{z}{p} / \tau\right)}{\sum{a \in \mathcal{A}{i}} \exp \left(\boldsymbol{z}{i} \cdot \boldsymbol{z}_{a} / \tau\right)}
\end{equation}

Despite its effectiveness, this approach still requires learning a linear classifier using the cross-entropy loss apart from the contrastive term. This is because the contrastive loss can only learn generic representations for the input examples. Thus, we argue that the supervised contrastive learning developed so far appears to be a naive adaptation of unsupervised contrastive learning to the classification

\section{Proposed Method} \label{sec:approach}

The keyword spotting task can be framed as a sequence classification problem, where the keyword spotting network maps an input audio sequence $X=\left\{x_{0}, x_{1}, \ldots, x_{T}\right\}$ to a set of keyword classes $Y \in y_{1: S}$. Here, $T$ represents the number of frames, and $S$ denotes the number of classes. Our proposed keyword spotting model, depicted in Fig~\ref{fig
}(A), consists of five key components: (1) Compressed Convolutional Layer, (2) Transformer Block, (3) Feature Selection Layer, (4) Bottleneck Layer, and (5) Projection Layer.

\subsection{Compressed Convolutional Layer}

The Compressed Convolutional Layer replaces the CNN block in the original design. This layer learns dense and informative frame representations from the input sequence $X$. Specifically, it utilizes convolutional neural networks (CNNs), an attention-based soft-pooling approach, and residual convolution blocks for feature extraction and compression.

\subsubsection{Frame Convolution}

Just as in the original CNN block, the convolution operation is applied to each frame. Given the input sequence $X$ and the $i$-th filter, the convolution for the $j$-th frame is expressed as

\begin{equation}
\mathbf{x}_{j}^{i}=\operatorname{conv}\left(\left\{\mathbf{x}_{j}, \mathbf{x}_{j+1}, \cdots, \mathbf{x}_{j+k_{i}-1}\right\} ; \mathbf{W}_{x}^{i}\right),
\end{equation}

where $\mathbf{W}_{x}^{i}$ is the learned parameter of the $i$-th filter.

\subsubsection{Attention-based Soft-pooling}

To eliminate redundant information in the speech dataset, we propose an attention-based soft-pooling operation on the frame representations learned by the previous equation. Specifically, given a frame $x_{j}$, its neighboring frames $\left\{x_{j+1}, \cdots, x_{j+g-1}\right\}$, and the corresponding filter $f_{i}$, we first learn the local-based attention scores $\alpha_{j}^{i}=\mathbf{W}_{\alpha}^{i} \mathbf{x}_{j}^{i}+b$ with softmax function, and then conduct the soft-pooling operation to obtain the compressed representation as in the following equation:

\begin{equation}
\mathbf{o}_{p}^{i}=\sum_{q=j}^{j+g-1} \beta_{q}^{i} \mathbf{x}_{q}^{i}
\end{equation}

\subsubsection{Residual Convolution Block}

We now have a denoised matrix $\left\{\mathbf{o}_{1}^{i}, \mathbf{o}_{2}^{i}, \cdots, \mathbf{o}_{P}^{i}\right\}$ that represents the input sequence $X$. To avoid vanishing gradients and facilitate model training, we introduce residual blocks on top of the compressed features. In particular, we replace the batch norm layer with the group norm layer. Let $a$ denotes the number of residual blocks, we have

\begin{equation}
\mathbf{r}_{p}^{i}=\operatorname{ResidualBlcok}\left(\left\{\mathbf{o}_{p}^{i}, \cdots, \mathbf{o}_{p+a-1}^{i}\right\}\right),
\end{equation}

where $\operatorname{ResidualBlock}$ is the operation of the residual convolution block. 
\subsection{ResLayer Block}
\subsubsection{Transformer Block}

The output from the Compressed Convolutional Layer, $R=\left\{\mathbf{r}_{1}^{i}, \mathbf{r}_{2}^{i}, \cdots, \mathbf{r}_{P}^{i}\right\}$, is then fed into the Transformer Block. This block captures long-term dependencies in the sequence via the self-attention mechanism: $E_{tran}=\text { Self-Attention }_{\times M}\left(R\right),$ where $M$ is the number of self-attention layers.

\subsubsection{Feature Selecting Layer}

Following the Transformer Block, the Feature Selecting Layer is implemented to extract keyword information from the sequence $E_{tran}$.

\begin{equation}
E_{\text {feat }}=\operatorname{Concat}\left(E_{\text {tran }}[T-r, T]\right),
\end{equation}

Here, the last $r$ frames of $E_{tran}$ are gathered, and all the collected frames are concatenated together into one feature vector $E_{feat}$.

\subsubsection{Bottleneck and Project Layers}

After the Feature Selecting Layer, a Bottleneck Layer and a Projection Layer are added. These layers map the hidden states to the predicted classification classes $\tilde{Y}$.

\begin{equation}
E_{bn}=\mathrm{FC}_{\mathrm{bn}}\left(E_{\text {feat }}\right),
\end{equation}

\begin{equation}
\tilde{Y}=\mathrm{FC}_{\operatorname{proj}}\left(E_{bn}\right),
\end{equation}

Finally, the cross-entropy (CE) loss for supervised learning and model fine-tuning is computed based on the predicted classes $\tilde{Y}$ and ground truth classes $Y$. $\mathcal{L}_{ce}=\operatorname{CE}(Y, \tilde{Y})$.


\begin{figure}
    \centering
    \includegraphics[width=0.76\linewidth]{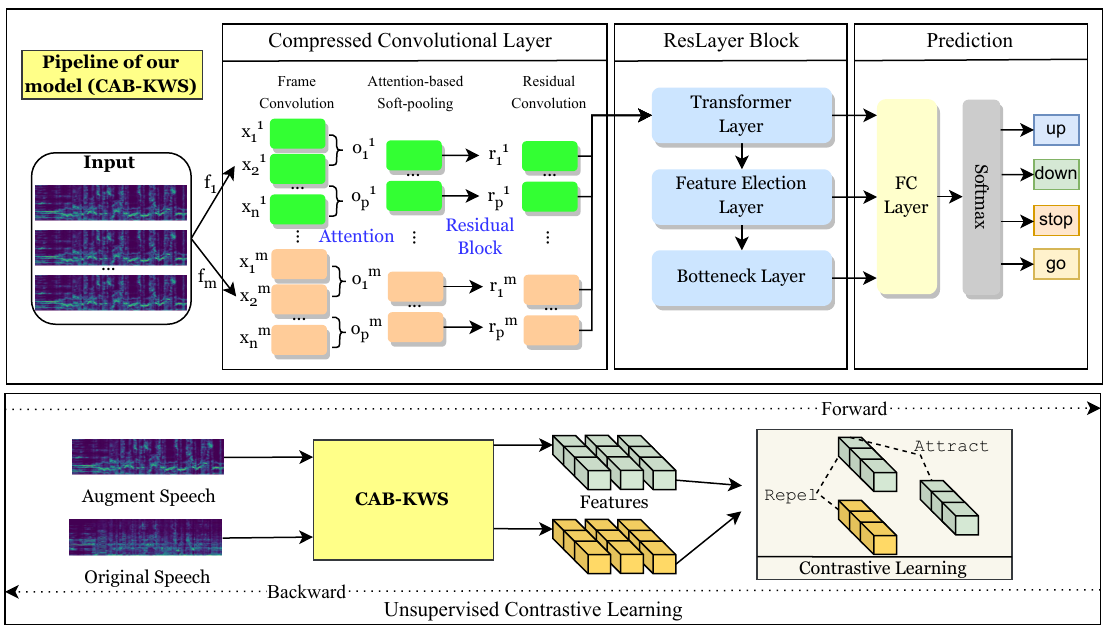}
    \caption{A: The architecture of our CAB-KWS for the keyword spotting task consists of a compressed layer, ResLayer Block, and Decision Block. B: The proposed method integrates speech augmentation with unsupervised and contrastive learning for audio processing.}
    \label{fig:overview}
\end{figure}

\subsection{Augmentation Method}
Data augmentation is a widely utilized technique to enhance model performance and robustness, particularly in speech-related tasks. In this study, we delve into speed and volume-based augmentation in the context of unsupervised learning for keyword detection. A specific audio sequence, represented as $X=A(t)$, is defined by its amplitude $A$ and time index $t$.

Regarding speed augmentation, a speed ratio symbolized by $\lambda_{\text {speed }}$ is established to modify the speed of $X$. The following formula describes this process:$X^{aug}=A\left(\lambda_{\text {speed }} t\right)$. For volume augmentation, similarly, we set an intensity ratio, $\lambda_{\text {volume }}$, to alter the volume of $X$, as presented in the following equation: $X^{aug}=\lambda_{\text {volume }} A(t)$. By using various ratios $\lambda_{\text {speed }}$ and $\lambda_{\text {volume }}$, we can generate multiple pairs of speech sequences, $\left(X, X^{a u g}\right)$, to facilitate the training of the audio representation network via unsupervised learning. The fundamental assumption is that speech utterances, regardless of speed or volume variations, should exhibit similar high-level feature representations for keyword-spotting tasks.






\subsection{Contrastive Learning Loss}
We aim to align the softmax transform of the dot product between the feature representation $\boldsymbol{z}{i}$ and the classifier $\boldsymbol{\theta}{i}$ of the input example $X_{i}$ with its corresponding label. Let $\boldsymbol{\theta}{i}^{*}$ denote the column of $\boldsymbol{\theta}{i}$ that corresponds to the ground-truth label of $\boldsymbol{x}{i}$. We aim to maximize the dot product $\boldsymbol{\theta}{i}^{* T} \boldsymbol{z}{i}$. To achieve this, we learn a better representation of $\boldsymbol{\theta}{i}$ and $\boldsymbol{z}_{i}$ using supervised signals.

The Dual Contrastive Loss exploits the relation between different training samples to maximize $\boldsymbol{\theta}{i}^{* T} \boldsymbol{z}{j}$ if $\boldsymbol{x}{j}$ has the same label as $\boldsymbol{x}{i}$, while minimizing $\boldsymbol{\theta}{i}^{* T} \boldsymbol{z}{j}$ if $\boldsymbol{x}{j}$ carries a different label from $\boldsymbol{x}{i}$.

To define the contrastive loss, given an anchor $\boldsymbol{z}{i}$ originating from the input example $\boldsymbol{x}{i}$, we take $\left\{\boldsymbol{\theta}_j^*\right\}_{j \in \mathcal{P}_i}$
as positive samples and $\left\{\boldsymbol{\theta}_j^*\right\}_{j \in \mathcal{A}_i \backslash \mathcal{P}_i}$ as negative samples. The contrastive loss is defined as follows:

\begin{equation}
\mathcal{L}{z}=\frac{1}{N} \sum{i \in \mathcal{I}} \frac{1}{\left|\mathcal{P}{i}\right|} \sum{p \in \mathcal{P}{i}}-\log \frac{\exp \left(\boldsymbol{\theta}{p}^{} \cdot \boldsymbol{z}{i} / \tau\right)}{\sum{a \in \mathcal{A}{i}} \exp \left(\boldsymbol{\theta}{a}^{} \cdot \boldsymbol{z}_{i} / \tau\right)}
\end{equation}

Here, $\tau \in \mathbb{R}^{+}$ is the temperature factor, $\mathcal{A}{i}:=\mathcal{I} \backslash{i}$ is the set of indexes of the contrastive samples, $\mathcal{P}_i := \left\{p \in \mathcal{A}_i : y_p = y_i\right\}$ is the set of indexes of positive samples, and $\left|\mathcal{P}{i}\right|$ is the cardinality of $\mathcal{P}{i}$. Similarly, given an anchor $\boldsymbol{\theta}{i}^{*}$, we take $\left\{\boldsymbol{z}_j\right\}_{j \in \mathcal{P}_i}$ as positive samples and $\left\{\boldsymbol{z}_j\right\}_{j \in \mathcal{A}_i \backslash \mathcal{P}_i}$ as negative samples. The contrastive loss is defined as follows:

\begin{equation}
\mathcal{L}_{\theta}=
\frac{1}{N} \sum_{i \in \mathcal{I}} \frac{1}{\left|\mathcal{P}{i}\right|} \sum{p \in \mathcal{P}{i}}-\log \frac{\exp \left(\boldsymbol{\theta}{i}^{} \cdot \boldsymbol{z}{p} / \tau\right)}{\sum{a \in \mathcal{A}{i}} \exp \left(\boldsymbol{\theta}{i}^{} \cdot \boldsymbol{z}_{a} / \tau\right)}
\end{equation}

Finally, Dual Contrastive Loss is the combination of the above two contrastive loss terms:

\begin{equation}
\mathcal{L}_{Dual}=\mathcal{L}{z}+\mathcal{L}_{\theta}
\end{equation}

As illustrated in Fig.~\ref{fig:overview}(B), the structure of the proposed unsupervised learning method rooted in augmentation, involves two primary steps akin to other unsupervised strategies: (1) unsupervised data undergoes initial pre-training and (2) supervised KWS data is then fine-tuned. The pre-training phase sees the extraction of a bottleneck feature by training the unlabelled speech, which is subsequently used for KWS prediction in the fine-tuning stage.

In pre-training, the paired speech data $\left(X, X^{a u g}\right)$ is fed into CNN-Attention models with identical parameters. Since $X^{a u g}$ is derived from $X$, the unsupervised method we've developed assumes that both $X$ and $X^{a u g}$ will yield analogous high-level bottleneck features. This implies the speech content remains identical regardless of the speaker's speed or volume. The network's optimization, therefore, must highlight the similarity between $X$ and $X^{a u g}$. The Mean Square Error (MSE) $\mathcal{L}_{s i m}$ is utilized to determine the distance between $X$ and $X^{a u g}$'s output.

\begin{equation}
\mathcal{L}{\text {sim }}=\frac{1}{U{b n}} \sum_{u=0}^{U_{b n}}\left|E_{b n}(u)-E_{b n}^{a u g}(u)\right|^{2}
\end{equation}

In this context, $U_{bn}$ represents the dimensions of the bottleneck feature vector, while $E_{bn}$ and $E_{bn}^{aug}$ correspond to the bottleneck layer outputs for the original speech $X$ and the augmented speech $X^{aug}$, respectively.

The network also includes an auxiliary training branch designed to predict the average feature of the speech segment input, helping the network learn the intrinsic characteristics of speech utterances. To achieve this, the average vector of the input Fbank vector $X$ is first calculated along the time axis $t$. A reconstruction layer connected to the bottleneck layer is then used to reconstruct this average Fbank vector $\tilde{X}$. The MSE loss $\mathcal{L}{x}$ is applied to measure the similarity between the original and reconstructed audio vectors along the feature dimension $U_{x}$.

\begin{equation}
\begin{aligned}
\mathcal{X} & =\frac{1}{T} \sum_{T}(X) \
\tilde{X} & =\mathrm{FC}{\text {reconstruct }}\left(E{b n}\right) \
\mathcal{L}{x} & =\frac{1}{U{x}} \sum_{u=0}^{U_{x}}|\mathcal{X}(u)-\tilde{X}(u)|^{2}
\end{aligned}
\end{equation}

In this context, $U_{x}$ denotes the dimension of the Fbank feature vector, and $\mathcal{X}$ represents the mean vector of $X$. The loss $\mathcal{L}_{x}^{\text{aug}}$ between the augmented average audio $\mathcal{X}^{\text{aug}}$ and the reconstructed feature $\tilde{X}^{\text{aug}}$ can be similarly defined as:

\begin{equation}
\mathcal{L}{x}^{a u g}=\frac{1}{U{x}} \sum_{u=0}^{U_{x}}\left|\mathcal{X}^{a u g}(u)-\tilde{X}^{a u g}(u)\right|^{2}
\end{equation}

Hence, the final unsupervised learning (UL) loss function $\mathcal{L}{u l}$ comprises of the three aforementioned losses $\mathcal{L}{\text {sim }}, \mathcal{L}{x}$, and $\mathcal{L}{x}^{a u g}$

\begin{equation}
    \mathcal{L}_{u l}=\lambda_{1} \mathcal{L}_{\text {sim }}+\lambda_{2} \mathcal{L}_{x}+\lambda_{3} \mathcal{L}_{x}^{a u g}+\lambda_{4} \mathcal{L}_{Dual}
\end{equation}
 
Where $\lambda_{1}, \lambda_{2}, \lambda_{3}, \lambda_{4}$ are the factor ratios of each loss component.

In the fine-tuning stage, the average feature prediction branch is discarded, and a projection layer, followed by a softmax layer, is added after the bottleneck layer for KWS prediction. The original network's parameters can either be kept fixed or adjusted during fine-tuning. Our experiments indicate that adjusting all parameters enhances performance, so we choose to update all parameters during this phase.

\section{EXPERIMENT SETUP}\label{sec:setup}
In this section, we evaluated the proposed method on keyword spotting tasks by implementing our CNN-Attention model with supervised training and comparing it to Google's model. An ablation study was conducted to examine the impact of speed and volume augmentation on unsupervised learning. Additionally, we compared our approach with other unsupervised learning methods, including CPC, APC, and MPC, using their published networks and hyperparameters without applying any additional experimental tricks [23]-[25]. We also analyzed how varying pre-training steps influence the performance and convergence of the downstream KWS task.

\subsection{Datasets}
We used Google's Speech Commands V2 Dataset~\citep{warden2018speech}for evaluating the proposed models. The dataset contains more than 100k utterances. Total 30 short words were recorded by thousands of different people, as well as background noise such as pink noise, white noise, and human-made sounds. The KWS task is to discriminate among 12 classes: "yes", "no”, "up”, "down", "left", "right”, "on", "off", "stop", "go", unknown, or silence. The dataset was split into training, validation, and test sets, with $80 \%$ training, $10 \%$ validation, and $10 \%$ test. This results in about 37000 samples for training, and 4600 each for validation and testing. We applied the HuNonspeech ${ }^{1}$ real noisy data to degrade the original speech. In our experiments, this strategy was executed using the Aurora4 tools $\left.\right|^{2}$. Each utterance was randomly corrupted by one of 100 different types of noise from the HuNonspeech dataset. The Signal Noise Ratio (SNR) for each utterance ranged from 0 to 20 dB, with an average SNR of $10 \mathrm{~dB}$ across all datasets.


\begin{table}[]
\centering
\caption{Results comparison of KWS Model, Classification Accuracy (\%)}
\begin{tabular}{llcc}
\toprule
Model Name & Supervised Training Data & Dev & Eval \\
\midrule
Sainath and Parada (Google) & Speech Commands & - & 84.7\\
CAB-KWS (w/o volume) & Speech Commands & 86.4 & 85.3 \\
CAB-KWS  \& speed augment & Speech Commands & \textbf{87.3} & \textbf{85.8} \\
\bottomrule
\end{tabular}
\end{table}

\begin{table}[]
\centering
\caption{Ablation study, the effect of speed and volume augmentation, classification accuracy (\%)}
\resizebox{0.8\linewidth}{!}{
\begin{tabular}{c|c|c|c|c}
\toprule
Model Name & Pre-training Data & Fine-tuning Data & Dev & Eval \\
\midrule
CAB-KWS + vo-pre. & Speech Commands & Speech Commands & 86.1 & 85.9 \\
\hline
CAB-KWS + sp-pre. & Speech Commands & Speech Commands & 87.8 & 86.9 \\
\hline
CAB-KWS + vo-sp-pre. & Speech Commands & Speech Commands & 87.9 & 87.2 \\
\hline
CAB-KWS + vo-sp-pre-contras. & Speech Commands & Speech Commands & \textbf{88.1} & \textbf{88.3 }\\
\midrule
CAB-KWS + vo-pre. & Librispeech-100 & Speech Commands & 86.3 & 86.0 \\
\hline
CAB-KWS + sp-pre. & Librispeech-100 & Speech Commands & 87.9 & 87.9 \\
\hline
CAB-KWS + vo-sp-pre. & Librispeech-100 & Speech Commands & 88.2 & 88.1 \\
\hline
CAB-KWS + vo-sp-pre-contras \& & Librispeech-100 & Speech Commands & \textbf{88.4} & \textbf{88.5} \\
\bottomrule
\end{tabular}}
\begin{tablenotes}
\footnotesize
    \item ``\textbf{vo-pre.}" means volume pre-training; ``\textbf{sp-pre.} " is speed pre-training; ``\textbf{vo-sp-pre.}" indicates volume \& speed pre-training; ``\textbf{contras.}"  is contrastive learning.
\end{tablenotes}
\label{tb:2}
\end{table}

As with other unsupervised approaches, a large unlabeled corpus, consisting of 100 hours of clean Librispeech~\citep{panayotov2015librispeech} audio, was used for network pre-training through unsupervised learning. Initially, the long utterances were divided into 1-second segments to align with the Speech Commands dataset. Following this, the clean segments were mixed with noisy HuNonspeech data using Aurora 4 tools, employing the same corruption mechanism as the Speech Commands.

\subsection{Model Setup}
The model architecture consists of:
\begin{itemize}
  \item CNN blocks with 2 layers, a 3x3 kernel size, 2x2 stride, and 32 channels.
  \item Transformer layer with 2 layers, a 320-dimensional embedding space, and 4 attention heads.
  \item Feature Selecting Layer retains the last 2 frames with a 2x320 dimension.
  \item Bottleneck Layer with a single fully connected (FC) layer of 800 dimensions.
  \item Project Layer with one FC layer outputting a 12-dimensional softmax.
  \item Reconstruct Layer with one FC layer outputting a 40-dimensional softmax.
\end{itemize}
The factor ratio is set to $\lambda_1=0.8$, $\lambda_2=0.05$, $\lambda_3=0.05$, and $\lambda_4=0.1$.

To demonstrate the effectiveness, we compared with other approaches:
\begin{itemize}
  \item Supervised Learning: Used Google's Sainath and Parada's model as baseline.
  \item Unsupervised Learning:
    \begin{itemize}
      \item Contrastive Predictive Coding (CPC): Learns representations via next step prediction.
      \item Autoregressive Predictive Coding (APC): Optimizes L1 loss between input and output sequences.
      \item Masked Predictive Coding (MPC): Utilizes Transformer with Masked Language Model (MLM) structure for predictive coding, incorporating dynamic masking.
    \end{itemize}
\end{itemize}

\section{EXPERIMENTAL RESULTS}\label{sec:results}

\subsection{Comparision of KWS Model (RQ1)}
The table compares the classification accuracy of three different KWS models: (1) the model by Sainath and Parada (Google), (2) the CAB-KWS model without volume augmentation, and (3) the CAB-KWS model with speed augment. It can be observed that the CAB-KWS model with speed augment achieved the highest classification accuracy on both the development (Dev) and evaluation (Eval) datasets. This research question aims to investigate how the inclusion of data augmentation techniques, specifically speed augment in this case, improves the performance of KWS models compared to models without these techniques. The results could be used to guide future development of KWS models and to optimize their performance for various applications.


\subsection{Ablation Study (RQ2)}
The CAB-KWS keyword spotting model is an advanced solution designed to improve the classification accuracy of speech recognition tasks. The ablation study presented in the table focuses on evaluating the impact of different pre-training techniques, such as volume pre-training, speed pre-training, combined volume and speed pre-training, and combined volume, speed, and contrastive learning pre-training, on the model's performance. By comparing the classification accuracy of CAB-KWS when fine-tuned on two datasets, Speech Commands and Librispeech-100, we can better understand the effectiveness of these pre-training techniques and their combinations.

Firstly, Tab.~\ref{tb:2} shows that the CAB-KWS model with speed pre-training (sp-pre.) outperforms the model with volume pre-training (vo-pre.) in both datasets. This result indicates that speed pre-training is more effective in enhancing the model's classification accuracy than volume pre-training. However, the combination of volume and speed pre-training (vo-sp-pre.) further improves the model's performance, demonstrating that utilizing both techniques can lead to better keyword spotting results.

Moreover, the inclusion of contrastive learning (contras.) in the pre-training process yields the highest classification accuracy in both Speech Commands and Librispeech-100 datasets. The CAB-KWS model with combined volume, speed, and contrastive learning pre-training (vo-sp-pre-contras.) outperforms all other models, highlighting the benefits of incorporating multiple pre-training methods. This result emphasizes the goodness of the CAB-KWS model, as it demonstrates its adaptability and capability to leverage various pre-training techniques to enhance its performance.

The CAB-KWS model's strength lies in its ability to capitalize on different pre-training methods, which can be tailored to suit specific datasets and tasks. By combining these techniques, the model can learn more robust and diverse representations of the data, leading to improved classification accuracy. This adaptability makes the CAB-KWS model particularly suitable for a wide range of applications in keyword spotting and speech recognition tasks, where performance and generalizability are of utmost importance.

In conclusion, the goodness of the CAB-KWS keyword spotting model is showcased through its ability to integrate various pre-training techniques, such as volume pre-training, speed pre-training, and contrastive learning, to improve classification accuracy. The ablation study demonstrates that the combination of these methods leads to the highest performance across different datasets, highlighting the model's adaptability and effectiveness in handling diverse keyword spotting tasks. This advanced model, with its robust pre-training methods and fine-tuning capabilities, offers a promising solution for speech recognition applications and can contribute significantly to advancements in the field.


\subsection{Comparison with Unsupervised Models (RQ3)}

\begin{table}[]
\centering  
\caption{Comparison results in accuracy (\%)}
\resizebox{\linewidth}{!}{
\begin{tabular}{c|c|c|c|c}
\toprule
Model Name & Pre-training Data & Fine-tuning Data & Dev &  Eval \\
\midrule
Contrastive Predictive Coding (CPC)  & Speech Commands & Speech Commands & 87.6 & 86.9 \\
\hline
Autoregressive Predictive Coding (APC)  & Speech Commands & Speech Commands & 87.2 & 86.5 \\
\hline
Masked Predictive Coding (MPC) & Speech Commands & Speech Commands & 87.0 & 86.7 \\
\hline
CAB-KWS (full) & Speech Commands & Speech Commands & \textbf{88.1} & \textbf{88.3} \\
\hline
Contrastive Predictive Coding (CPC)  & Librispeech-100 & Speech Commands & 87.8 & 87.4 \\
\hline
Autoregressive Predictive Coding (APC)  & Librispeech-100 & Speech Commands & 87.7 & 87.5 \\
\hline
Masked Predictive Coding (MPC)  & Librispeech- 100 & Speech Commands & 87.9 & 87.0 \\
\hline
CAB-KWS(full) & Librispeech-100 & Speech Commands & \textbf{88.4} & \textbf{88.5} \\
\bottomrule
\end{tabular}}
\label{tb:3}
\end{table}

The CAB-KWS model is a sophisticated keyword spotting solution that integrates multiple pre-training techniques to improve classification accuracy in speech recognition tasks. The Tab.~\ref{tb:3} provided presents a comparison of the CAB-KWS model with three other models that employ individual pre-training methods, namely Contrastive Predictive Coding (CPC), Autoregressive Predictive Coding (APC), and Masked Predictive Coding (MPC). By comparing the performance of these models, we can gain insights into the effectiveness of the CAB-KWS model and highlight its advantages over models based on single pre-training techniques.

The comparison in the table reveals that the CAB-KWS model consistently achieves the highest classification accuracy on both the development (Dev) and evaluation (Eval) datasets when fine-tuned on Speech Commands, regardless of the pre-training data source (Speech Commands or Librispeech-100). This result underlines the goodness of the CAB-KWS model as it demonstrates its ability to effectively utilize multiple pre-training techniques to outperform models that rely on individual pre-training methods.

The CAB-KWS model's superior performance can be attributed to its ability to integrate and capitalize on the strengths of various pre-training techniques. By combining different methods, the model can learn more diverse and robust representations of the data, which in turn leads to improved classification accuracy. This adaptability makes the CAB-KWS model particularly suitable for a wide range of applications in keyword spotting and speech recognition tasks, where performance and generalizability are crucial.

Furthermore, the CAB-KWS model's consistent performance across different pre-training data sources indicates its flexibility and robustness. It is not limited by the choice of pre-training dataset, which is an essential aspect of its goodness. This characteristic allows the model to be adaptable and versatile, enabling its use in various speech recognition applications with different data sources.

In summary, the CAB-KWS keyword spotting model showcases its goodness by effectively combining multiple pre-training techniques to achieve superior classification accuracy compared to models based on individual pre-training methods. Its consistent performance across different pre-training data sources highlights its adaptability, making it a promising solution for diverse speech recognition tasks. The CAB-KWS model's ability to harness the strengths of various pre-training techniques and deliver enhanced performance demonstrates its potential to contribute significantly to advancements in the field of speech recognition.





\begin{figure}[h]
\centering
\includegraphics[width=0.7\linewidth]{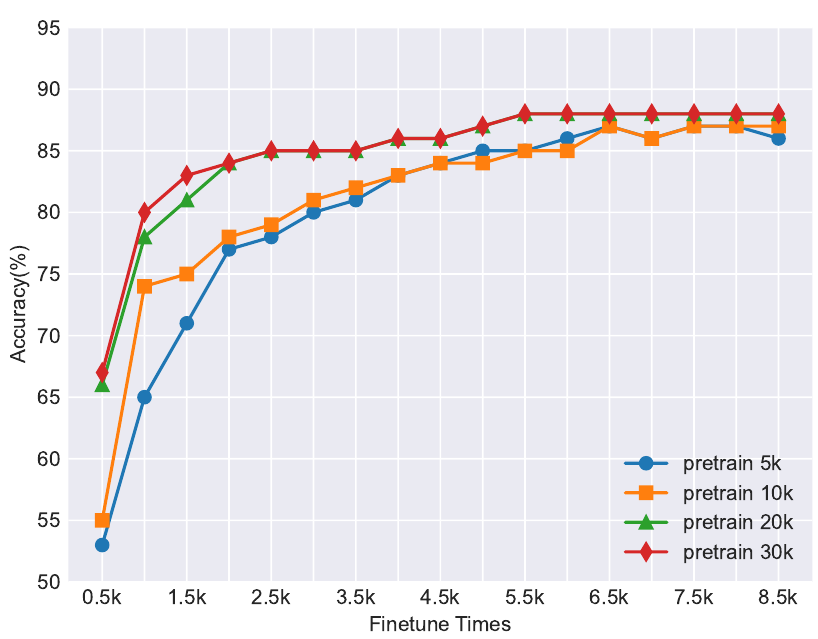}
\caption{Comparison of results with different pre-training steps. The number of pre-training steps in unsupervised learning significantly impacts accuracy and fine-tuning convergence. In our experiments, pre-training for $30 K$ steps achieved the best classification accuracy and the quickest convergence.}
\label{fig:finetune}
\end{figure}




\section{CONCLUSION}\label{sec:conclusion}
This paper presents a robust approach for the Keyword Spotting (KWS) task. Our CNN-Attention architecture, in combination with our unsupervised contrastive learning method, CABKS, utilizes unlabeled data efficiently. This circumvents the challenge of acquiring ample labeled training data, particularly beneficial when target keywords change or when positive samples are scarce. Furthermore, our speech augmentation strategy enhances the model's robustness, adapting to variations in keyword utterances. By using contrastive loss within mini-batches, we've improved training efficiency and overall performance. Our method outperformed others such as CPC, APC, and MPC in experiments. Future work could explore this approach's application to other speech tasks and investigate other augmentations or architectures to enhance performance. This work marks a significant step towards more reliable voice-controlled systems and interactive agents.

\bibliographystyle{splncs04}
\bibliography{ref.bib}
\end{document}